\documentclass[aps,prl,twocolumn,superscriptaddress]{revtex4-1}
\pdfoutput=1
\usepackage{mathrsfs}
\usepackage{graphicx}
\usepackage{amsmath}
\usepackage{amssymb}
\usepackage{color}
\usepackage{bm}

\definecolor{red}{rgb}{0.75,0,0}
\definecolor{blue}{rgb}{0,0,0.75}
\definecolor{green}{rgb}{0,0.5,0}

\newcommand{\vdw}[0]{vdW}
\newcommand{\x}[0]{{\bf\hat{x}}}
\newcommand{\y}[0]{{\bf\hat{y}}}
\newcommand{\X}[0]{{\bf\hat{X}}}
\newcommand{\Y}[0]{{\bf\hat{Y}}}
\newcommand{\Z}[0]{{\bf\hat{Z}}}

\begin{document}
\title{Molecular Tilt on Monolayer-Protected Nanoparticles}

\author{L. Giomi}
\affiliation{School of Engineering and Applied Sciences, Harvard University, Cambridge, Massachusetts 02138, USA}

\author{M. J. Bowick}
\affiliation{Department of Physics, Syracuse University, Syracuse, New York 13244, USA}

\author{X. Ma}
\affiliation{Department of Physics, Syracuse University, Syracuse, New York 13244, USA}

\author{A. Majumdar}
\affiliation{Oxford Centre for Collaborative Applied Mathematics, University of Oxford, OX1 3LB, UK}

\date{\today}

\begin{abstract}
The structure of the tilted phase of monolayer-protected nanoparticles is investigated by means of a simple Ginzburg-Landau model. The theory contains two dimensionless parameters representing the preferential tilt angle and the ratio $\epsilon$ between the energy cost due to spatial variations in the tilt of the coating molecules and that of the van der Waals interactions which favors uniform tilt. We analyze the model for both spherical and octahedral particles. On spherical particles, we find a transition from a tilted phase, at small $\epsilon$, to a phase where the molecules spontaneously align along the surface normal and tilt disappears. Octahedral particles have an additional phase at small $\epsilon$ characterized by the presence of six topological defects. These defective configurations provide preferred sites for the chemical functionalization of monolayer-protected nanoparticles via place-exchange reactions and their consequent linking to form molecules and bulk materials.
\end{abstract}

\maketitle

{\bf Introduction.} $-$ Monolayer-protected nanoparticles (MPNPs) are small aggregates of metallically-bonded atoms 2-8 nanometers in diameter, coated with organic ligands. Organic materials readily absorb on surfaces of metal and metal oxides to lower the interfacial energy between the substrate and the environment. These adsorbates naturally alter the surface properties and can have a significant influence on the stability of nanostructures \cite{Daniel:2004,Love:2005}. MPNPs have received considerable attention in the past few years due to the unique chemical and physical properties that make them promising candidates in a wide array of potential applications, including electronics \cite{andres:1996,collier:1997}, nanomedicine \cite{han:2007,pasquato:2005,georganopoulou:2005}, optics \cite{Ditlbacher:2002}, chemical microfabrication \cite{Mirkin:2011} and biology \cite{Salata:2004}.

\begin{figure}[b]
\centering
\includegraphics[width=1\columnwidth]{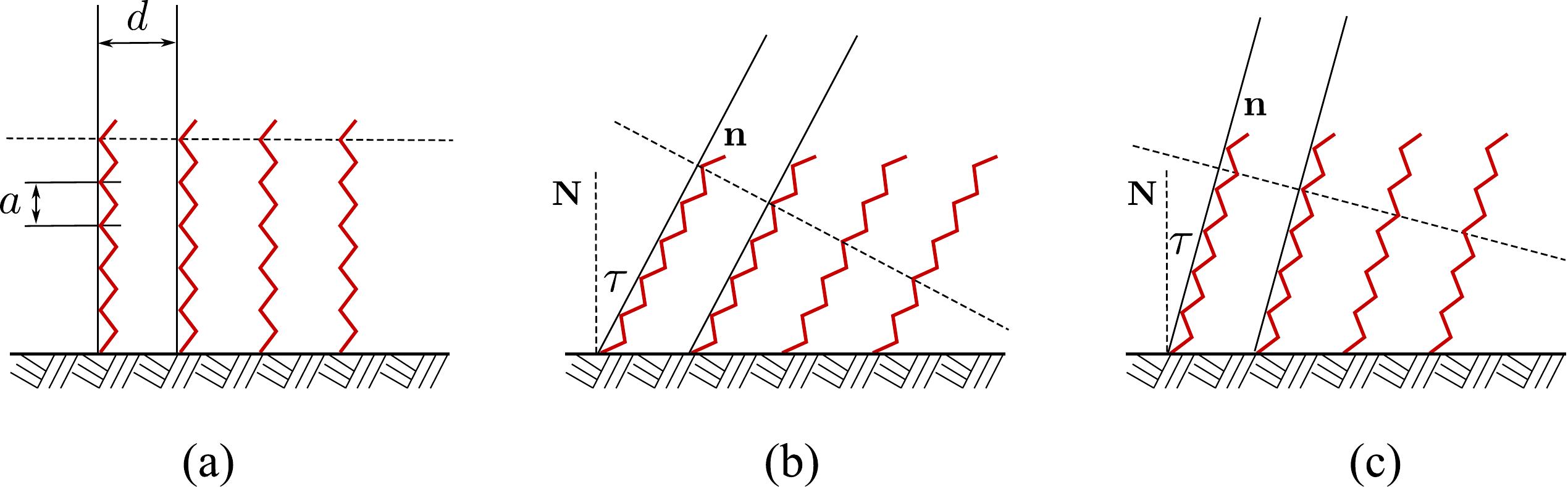}
\caption{\label{fig:tails}Schematic representation of thiol-terminated hydrocarbon chains attached to a substrate. (a) Upright configuration, (b) an optimal tilt and (c) a non-optimal tilt.}
\end{figure}

Gold nanoparticles coated with alkanethiol chains, e.g. CH$_{3}$(CH$_{2}$)$_{m}$SH, form one of the most thoroughly investigated classes of MPNPs because of their ease of fabrication. Due to the strong interaction between alkanethiols and the gold substrate, the sulfur head groups of thiols chemisorb on specific sites forming a commensurate lattice on the substrate. At low surface coverage the two free lone pair electrons of sulfur drive the alkyl chain close to the gold surface (lying down phase). As the coverage increases van der Waals (\vdw{}) interactions between the alkyl chains become important and the resultant competition leads to a configuration with the molecules standing upright on the substrate, but tilted with respect to the substrate normal. The minimum energy packing is achieved with the chains tilted at specific angles dictated by interlocking of the zigzag skeletal structure of the carbon atoms (Fig. \ref{fig:tails}).

\begin{figure}[t]
\centering
\includegraphics[width=.9\columnwidth]{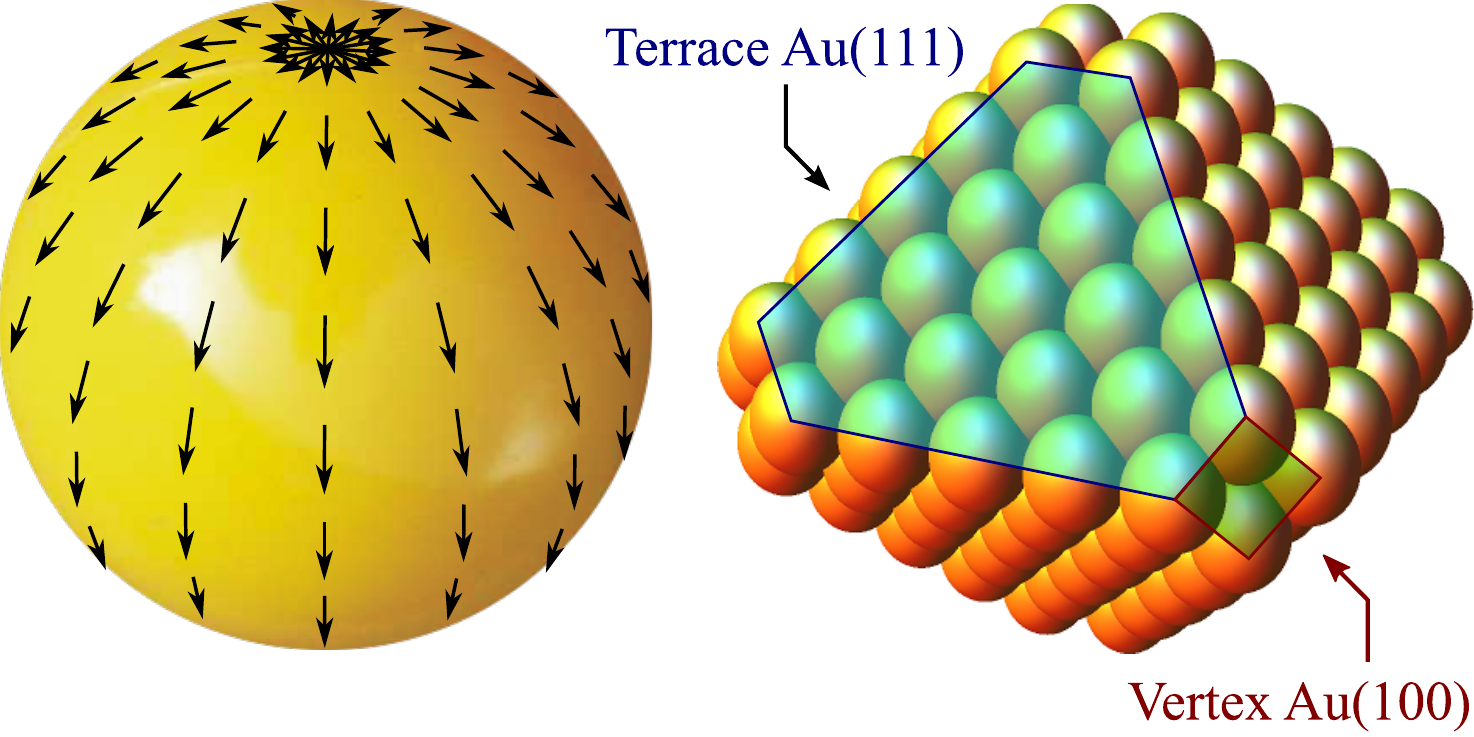}
\caption{\label{fig:textures}(Left) Bipolar texture with a source and a sink at the north and south pole. (Right) A model of a gold NP with octahedral geometry.}
\end{figure}

The morphology of a monolayer of tilted alkanethiols on the faceted surface of a gold nanoparticle has been the subject of intense investigation for over ten years \cite{Landman:1998,Zanchet:2000,Landman:2004,Ghorai:2007}. By means of molecular dynamics simulations it was found that, as a consequence of the seamlessness of the underlying substrate, the alkane chains tend to tilt uniformly across the nanoparticle, rather than assuming their equilibrium tilt angle on each facet, which would generate an energetically costly mismatch at the edges. On a closed surface such as a sphere or a polyhedron, however, perfectly uniform tilt can never be achieved. The projections of the tilted alkane chains on the tangent plane of the nanoparticle form a two-dimensional vector field which, by virtue of the Poincar\'e-Hopf theorem, must vanish in some isolated points \cite{Eisenberg:1979,Bowick:2009,Turner:2010}. This topological constraint translates into the proverbial statement that it is impossible to ``comb a hairy ball without creating a cowlick''. More precisely, one can introduce the notion of topological charge $q$ of a defect in a vector field as the angle (modulo $2\pi$) the vector field rotates in one counter-clockwise circuit of any closed contour enclosing the defect. Thus a source and a sink have $q=1$, while a saddle point has $q=-1$. The Poincar\'e-Hopf theorem requires that the sum of the topological charge of all the isolated defects (zeros) of a vector field on an oriented differentiable manifold $M$ is equal to the Euler characteristic $\chi$ of $M$. On a sphere, for which $\chi=2$,  any vector field must then contain defects, such as a source and a sink each with topological charge one (Fig. \ref{fig:textures}, left).

The inevitable presence of topological defects on MPNPs has stimulated the imagination of theorists and experimentalists who envisioned the possibility of exploiting functionalized defects as preferential binding sites of a nanoparticle. In this way one can engineer an entire warehouse of ``meso-atoms'' of controlled valence and directionality of bonds \cite{Nelson:2002}. This program has been successfully implemented in the beautiful work of DeVries \emph{et al}. \cite{DeVries:2007} in which gold nanoparticles are coated with two species of ligands of different length which then undergo microphase separation. The defects can then be functionalized by chemically attaching sulfur terminated chains, such as 11-mercaptoundecanoic acid (MUA). This results in a divalent gold nanoparticle with directional bonds 180$^{\circ}$ apart. These in turn can be linked to create polymers and free-standing films. The mechanism that allows the binding of functional groups to the monolayer coating a nanoparticle is referred to as a place-exchange reaction and occurs when a thiol-terminated molecule on the nanoparticle detaches from the monolayer and is replaced by another molecule from the surrounding solvent. As observed in Refs. \cite{DeVries:2007,Zerbetto:2007} such a reaction is more likely to occur at a spot of the monolayer where a molecule is less tightly bound. A defective site is clearly the most natural candidate for  such a ``weak spot'', since the tilt angle differs here from that favored by \vdw\ interactions.

The fundamental degree of freedom to describe molecular tilt on a two-dimensional substrate is the \emph{tilt vector} \cite{MacKintosh:1991} corresponding to the projection of the molecular axis ${\bf n}$ on the plane of the substrate:
\begin{equation}\label{eq:tilt-vector}
{\bf t} = \frac{{\bf n}}{{\bf n}\cdot{\bf N}}-{\bf N}
\end{equation}
with ${\bf N}$ the substrate normal vector (see Fig. \ref{fig:tails}). The normalization is chosen in such a way that $t=|{\bf t}|=\tan\tau$, with $\tau$ the tilt angle. The interlocking condition between chains translates into the requirement $t=t_{0}=na/d$,  with $a$ the distance between carbon atoms in a chain and $d$ the spacing between neighboring chains \cite{Outka:1987}. A simple description of molecular tilt on coated nanoparticles can be achieved by introducing a Ginzburg-Landau (GL) model with the following energy functional:
\begin{equation}\label{eq:energy1}
E = \int dA\,\left[\tfrac{1}{2}K_{A}\nabla_{i}t_{j}\nabla^{i}t^{j}+\tfrac{1}{4}\lambda(t^{2}-t_{0}^{2})^{2}\right],
\end{equation}
where $\nabla_{i}$ is the covariant derivative along the surface of the substrate. The first term in Eq. \eqref{eq:energy1} favors uniform tilt, while the second term implies that molecules have a preferential tilt angle $\tau_{0}=\arctan t_{0}$, typically in the range $30^{\circ}$- $45^{\circ}$ \cite{sellers:1993,porter:1987,nuzzo:1990,ehler:1997}. The parameter $\lambda$ is directly related to the strength of the \vdw\ interactions between carbons. The physically observable states may be described as minimizers of the above GL energy. In the following we will identify stable equilibrium configurations ${\bf t}$ for the energy \eqref{eq:energy1} for the case of a spherical substrate and an octahedral substrate. The spherical substrate approximates the shape of a large NP, while the octahedral substrate is appropriate for small NPs where the faceted structure of the particle is noticeable (Fig. \ref{fig:textures}, right).

{\bf Spherical nanoparticles.} $-$ On a sphere, the tilt vector can be conveniently expressed in a local orthonormal frame $\{{\bf e}_{1},{\bf e}_{2}\}$, namely ${\bf t}=t(\cos\alpha\,{\bf
e}_{1}+\sin\alpha\,{\bf e}_{2})$. The gradient terms can then be expressed as:
\begin{equation}
\nabla_{i}t_{j}\nabla^{i}t^{j} = g^{ij}\partial_{i}t\,\partial_{j}t + t^{2}g^{ij}(\partial_{i}\alpha-A_{i})(\partial_{j}\alpha-A_{j}),
\end{equation}
where $g^{ij}$ is the metric tensor and $A_{i}={\bf e}_{1}\cdot\,\partial_{i}{\bf e}_{2}$ is the spin connection, whose curl is the Gaussian curvature of the surface: $K=\epsilon^{ij}\nabla_{i}A_{j}$, with $\epsilon^{ij}$ the Levi-Civita tensor \cite{Bowick:2009,Turner:2010}. We compute the Euler-Lagrange equations associated with \eqref{eq:energy1} to find that for a stable equilibrium tilt vector field, $t$ and $\alpha$ satisfy the following coupled partial differential equations:
\begin{subequations}\label{eq:landau-ginzburg1}
\begin{gather}
\Delta t + \left(\frac{\lambda}{K_{A}}\,t_{0}^{2}-|\nabla\alpha-{\bf A}|^{2}\right)t-\frac{\lambda}{K_{A}}\,t^{3}=0\\[7pt]
\nabla\cdot\left[t^{2}(\nabla\alpha-{\bf A})\right]=0.
\end{gather}
\end{subequations}
These equations are formally identical to the GL equations for superconductors \cite{Tinkham} with the $t$ playing the role of the absolute value of the wave function, $\alpha$ its phase and ${\bf A}$ the vector potential associated with an externally applied magnetic field. Furthermore the quantity $t^{2}(\nabla\alpha-{\bf A})$ plays the role of the current density and Eq. (\ref{eq:landau-ginzburg1}b) is equivalent to the charge conservation condition.

As mentioned in the introduction, the simplest configuration formed by tilted molecules on a spherical particle consists of two topological defects, a source and a sink, placed at the two poles of the sphere (Fig. \ref{fig:textures}, left). Using standard spherical coordinates $\theta$ and $\phi$, it is easy to see that, for this \emph{bipolar} textures, $\alpha$ is constant and ${\bf A}=A_{\phi}\bm{\hat{\phi}}$ with $A_{\phi}=-\cos\theta$. Eq. (\ref{eq:landau-ginzburg1}b) is then automatically satisfied provided $t$ is azimuthally symmetric as one would expect:
\begin{equation}
\nabla\cdot\left[t^{2}(\nabla\alpha-{\bf A})\right]=-2tA_{\phi}\nabla_{\phi}t = 0.
\end{equation}
Then, introducing the normalized tilt vector ${\bf u}={\bf t}/t_{0}$, Eq. (\ref{eq:landau-ginzburg1}a) can be written as
\begin{equation}\label{eq:gl-bipolar}
\partial_{\theta}^{2}u+\cot\theta\,\partial_{\theta}u+(2/\epsilon^{2}-\cot^{2}\theta)u-(2/\epsilon^{2})\,u^{3} = 0,
\end{equation}
where $u=|{\bf u}|$ and $\epsilon^{2}=2K_{A}/\lambda t_{0}^{2}R^{2}$ is a dimensionless parameter representing the ratio between the energy cost  due to spatial variations in the tilt of the chains, proportional to $K_{A}/R^{2}$ (where $R$ is the sphere radius) and that of the \vdw\ interactions which favors a uniform tilt and is proportional to $\lambda t_{0}^2$.

\begin{figure}[t]
\centering
\includegraphics[width=1\columnwidth]{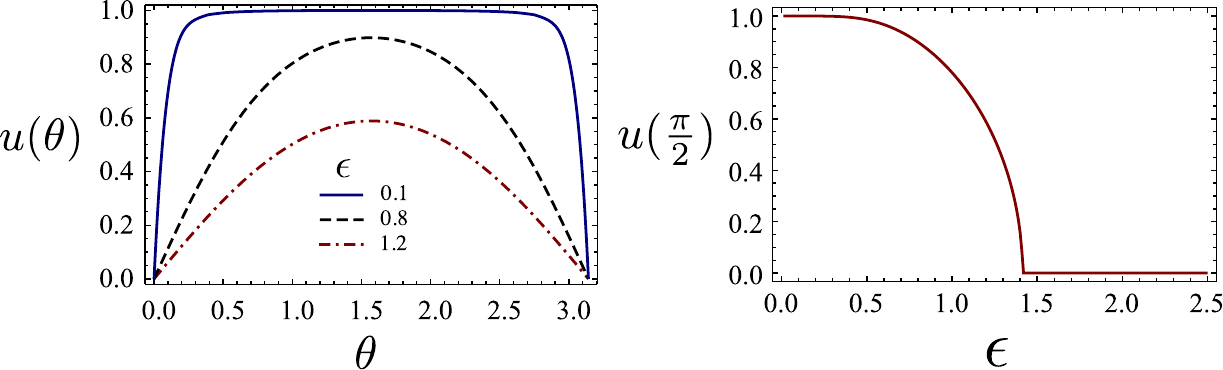}
\caption{\label{fig:tilt-angle}(Left) The magnitude of the normalized tilt vector $u=|{\bf t}|/t_{0}$ along a meridian for a bipolar texture for different values of $\epsilon$. (Right) The magnitude of the normalized tilt vector along the equator as a function of $\epsilon$ for the same system. The plot shows a transition at $\epsilon=\sqrt{2}$.}
\end{figure}

We display the numerical solution of Eq. \eqref{eq:gl-bipolar} for different values of $\epsilon$  in Fig. \ref{fig:tilt-angle}. For small $\epsilon$ the magnitude of the tilt vector increases rapidly from zero, at the defects, to the preferential value $t_{0}$. On the other hand, for $\epsilon>\sqrt{2}$, the tilted solution is no longer stable and the system undergoes a transition to a state where all the molecules are parallel to the surface normal, and thus $t=0$. This scenario is equivalent to the transition to a superconducting phase with the curvature of the substrate playing the role of an external magnetic field. Just as the magnetic field is given by the curl of the vector potential, the Gaussian curvature is given by the curl of the spin connection. The bipolar texture is equivalent to the mixed state in which the magnetic field penetrates the sample in only a limited number of defective spots. Increasing the magnetic field eventually destroys the superconducting state and the sample is now completely permeable. Analogously, at small curvatures (large radii) the chains are uniformly tilted with the exception of the two topologically required defects at the north and south pole. When the curvature is increased above a critical value, the system undergoes a transition to a state where $t=0$ everywhere or, equivalently, every point is defective.

{\bf Octahedral nanoparticles.} $-$ Small nanoparticles are far from spherical in shape. In this case we might expect the faceted structure to favor the formation of more complicated textures that match more effectively the geometry of the underlying substrate. Here we consider a simple approximation of the actual shape consisting of a regular octahedron of edge-length $\ell$. The stable equilibria $\mathbf{u}$ satisfy the classic GL-equations on each triangular face $T_{i}$ ($i=1,\,2,\,\cdots \,8$):
\begin{subequations}\label{eq:landau-ginzburg2}
\begin{align}
&\Delta {\bf u} +\frac{2}{\epsilon}\,(1-u^{2})\,{\bf u}  = 0\,, \qquad {\bf r} \in T_{i}\,,\\[5pt]
&{\bf u}  = \bm{\sigma}({\bf r})\,, \qquad  {\bf r} \in \partial T_{i}\,,
\end{align}
\end{subequations}
where ${\bf r}$ the position vector and $\bm{\sigma}$ is an appropriate function on the boundary $\partial T_{i}$ of the face $T_{i}$. Calculating the energy of a given configuration then involves prescribing a suitable function $\bm{\sigma}$ and solving the boundary-value problem defined by Eqs. \eqref{eq:landau-ginzburg2}. We will in fact consider three different boundary-value problems, corresponding to different physical scenarios, and prescribe a suitable form of $\bm{\sigma}$ in each case. {\em 1)} The chains rotate smoothly across the edge joining two neighboring faces. {\em 2)} The chains keep their preferential tilt angle along the perimeter of each face. We impose tangential boundary conditions, which require the tilt vector to be parallel to the edges. This necessarily creates discontinuities at the triangular vertices. {\em 3)} An intermediate configuration where both tangent and continuous boundary conditions coexist. Both the first and second scenarios have been discussed in the literature, where they are sometimes referred to as {\em continuous} \cite{Landman:2004} and {\em crystallographic} \cite{Zanchet:2000} models, and using molecular dynamics simulation it has been found that they are both possible, depending on the length of the chains as well as the temperature of the system~\cite{Ghorai:2007}.

\begin{figure}
\centering
\includegraphics[width=0.8\columnwidth]{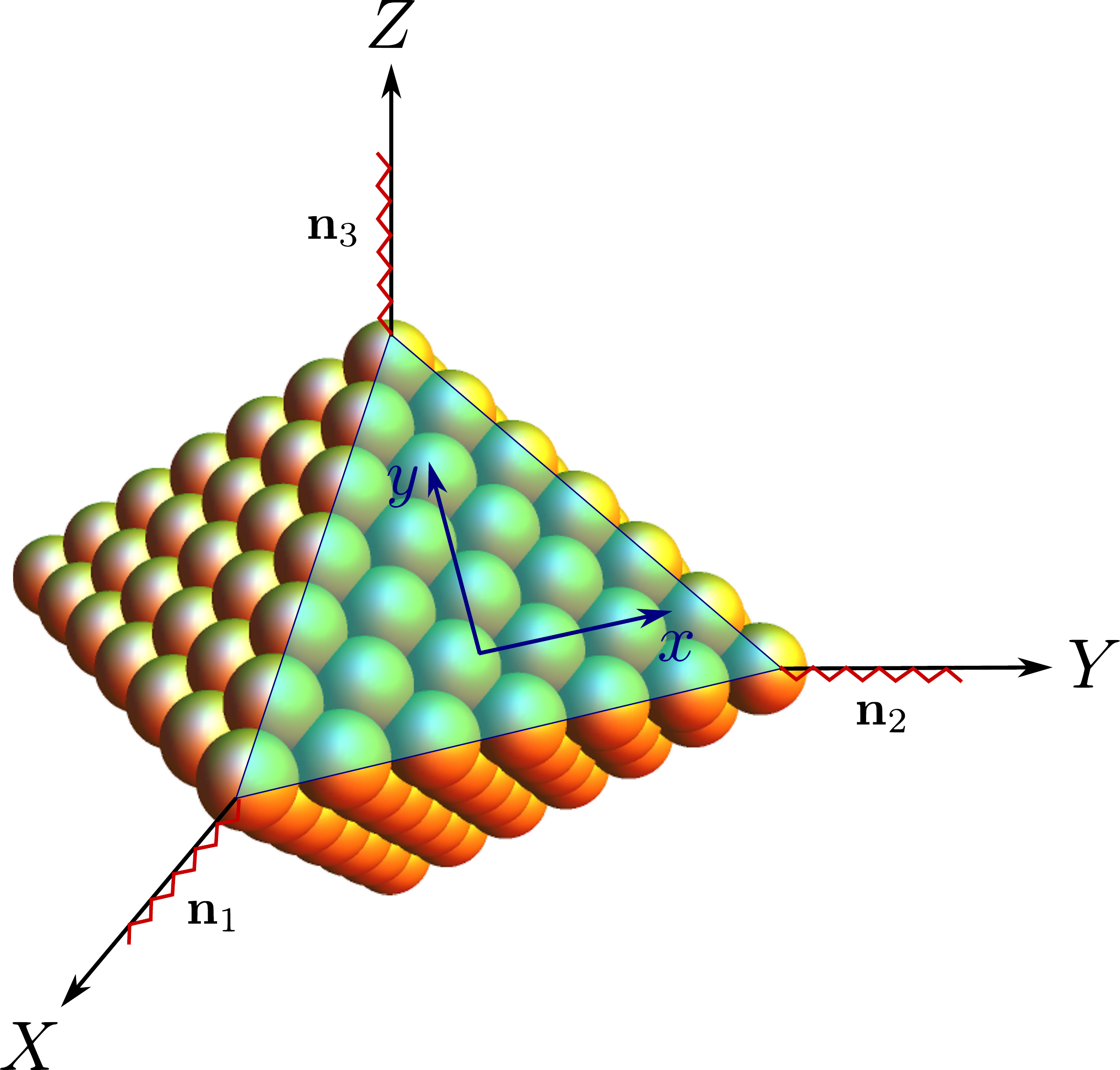}
\caption{\label{fig:octahedron}Schematic of an octahedral nanoparticle with the chains oriented at the vertices of a single triangular face as prescribed by the continuous model.}
\end{figure}

\begin{figure}
\centering
\includegraphics[width=0.9\columnwidth]{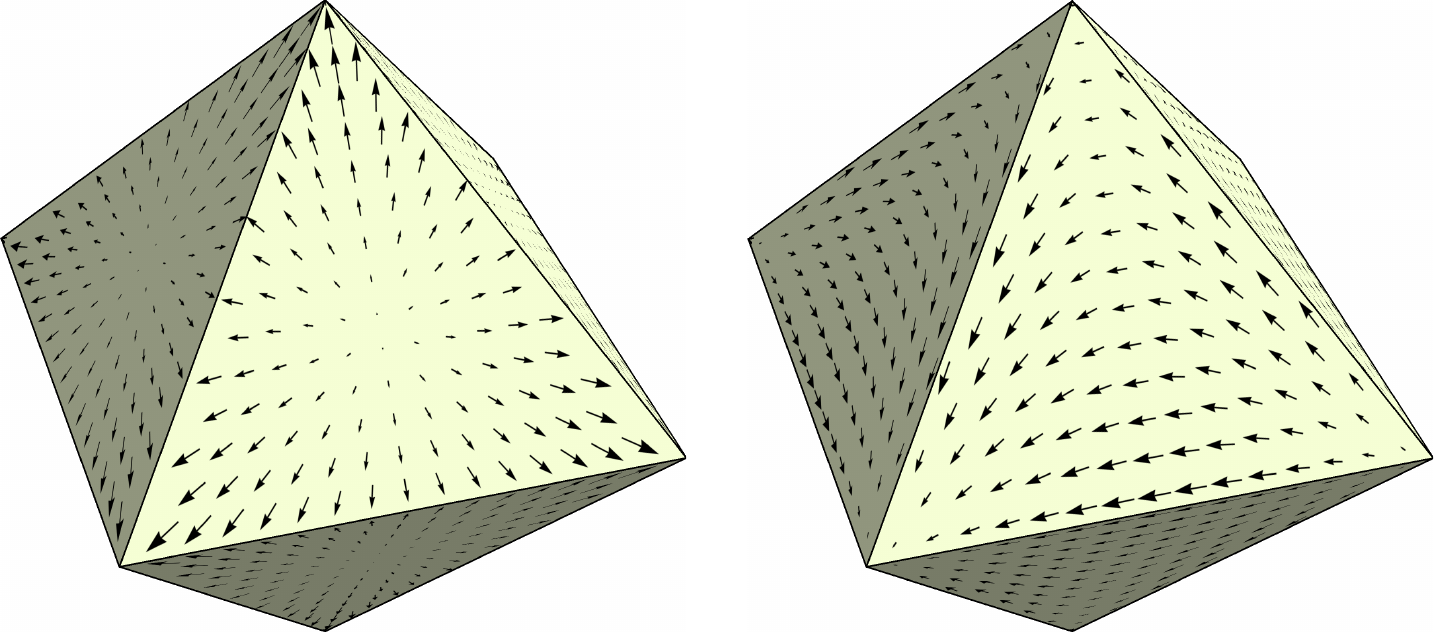}
\caption{\label{fig:textures2}The tilt vector for the continuous and crystallographic models. (Left) Continuous model; the chains smoothly rotate across the edges and vertices. This gives rise to a highly defective configuration consisting of 8 sources, 6 sinks and 12 saddles. (Right) Crystallographic model; each triangular face contains a source, a saddle and a sink, located at the vertices. The resulting configuration has 6 defects.}
\end{figure}

In the following we will see how these three scenarios can be implemented in our framework by a suitable choice of $\bm{\sigma}$ for Eqs. \eqref{eq:landau-ginzburg2}.  The first situation is modeled by requiring the chains at the six vertices of the octahedron to be parallel to the bisector of the solid angle subtended by each vertex (Fig. \ref{fig:octahedron}). Thus looking at the face in the positive octant of the lab frame $\{\X,\,\Y,\,\Z\}$, the chain orientations (molecular axes) at the three vertices are given by
\[
{\bf n}_{1} = \X\,,\quad
{\bf n}_{2} = \Y\,,\qquad
{\bf n}_{3} = \Z\,.
\]
The normal vector to the face is given by:
\begin{equation}\label{eq:normal-vector}
{\bf N}=(\X+\Y+\Z)/\sqrt{3}\,.
\end{equation}
In order to construct the function $\bm{\sigma}$, we introduce a smooth interpolation between the three vectors ${\bf n}_{1}$, ${\bf n}_{2}$ and ${\bf n}_{3}$. The simplest possible choice is a linear interpolation as shown below:
\begin{align*}
{\bf n}_{12}(s) &= (1-s)\X+s\Y\,,\\[5pt]
{\bf n}_{23}(s) &= (1-s)\Y+s\Z\,,\\[5pt]
{\bf n}_{31}(s) &= (1-s)\Z+s\X\,,
\end{align*}
where $s\in[0,1]$ is the normalized arc-length variable along each edge, measured counterclockwise, so that ${\bf n}_{12}(0)={\bf n}_{1}$, ${\bf n}_{12}(1)={\bf n}_{2}$ and similarly for the vector fields ${\bf n}_{23}(s)$ and ${\bf n}_{31}(s)$. The corresponding tilt vector can be readily calculated using Eq. \eqref{eq:tilt-vector}. Expressing this in the local ${\x,\,\y}$ frame:
\[
{\bf\hat{x}} = \sqrt{\tfrac{1}{2}}\,({\bf\hat{Y}}-{\bf\hat{X}})\,,\qquad
{\bf\hat{y}} = \sqrt{\tfrac{2}{3}}\,({\bf\hat{Z}}-\tfrac{1}{2}\,{\bf\hat{X}}-\tfrac{1}{2}\,{\bf\hat{Y}})\,,
\]
shows that the function $\bm{\sigma}$ is given by
\begin{align*}
\bm{\sigma}_{12}(s) &= t_{0}^{-1}\left[\sqrt{\tfrac{3}{2}}\,(2s-1)\,\x-\sqrt{\tfrac{1}{2}}\,\y\right],\\[5pt]
\bm{\sigma}_{23}(s) &= t_{0}^{-1}\left[-\sqrt{\tfrac{3}{2}}\,(s-1)\,\x+\sqrt{\tfrac{1}{2}}\,(3s-1)\,\y\right], \\[5pt]
\bm{\sigma}_{31}(s) &= t_{0}^{-1}\left[-\sqrt{\tfrac{3}{2}}\,s\,\x-\sqrt{\tfrac{1}{2}}\,(3s-2)\,\y\right].
\end{align*}
The magnitude of $\bm{\sigma}$ is the same on each edge: $\sigma=\sqrt{2+6s(s-1)}/t_{0}$. At the vertices, the tilt angle is thus given by $\tau_{\rm vertex}=\arctan\sqrt{2}\approx 55^{\circ}$. This depends exclusively on the geometry of the octahedron and implies that any tilted phase whose preferential angle differs from $55^{\circ}$ will be {\em frustrated}, in the sense that the preferential local order cannot propagate everywhere in the system. Other interpolations are also possible, but those do not change the overall picture. This configuration is highly defective. Because of the natural splay introduced by the orientation of the tilt vector at the vertices, each triangular face has a source at the centroid, a sink at each vertex and a saddle at the middle of each edge (Fig. \ref{fig:textures2}, left). Hence, there are a total of 26 defects. Calculating the total topological charge of this configuration provides a realization of the Euler characteristic of convex polyhedra: $Q_{\text{tot}}=V-E+F=2$, where $V$, $E$ and $F$ are respectively the number of vertices, edges and faces of the octahedron and the sign is that of the corresponding topological charge.

For the crystallographic model we impose tangential boundary conditions on all three edges. This means that on a triangular edge, $\mathbf{t}$ has to be parallel to that edge. This necessarily creates defects at the vertices: each triangle has exactly one source, one saddle and one sink located at the vertices. The magnitude $t$ of the tilt vector will be approximately constant and equal to the preferential value $t_{0}$ except for a thin boundary layer near each vertex where $t=0$. A suitable functional form is given by $t = t_{0}\tanh s(1-s)/\epsilon$. The function $\sigma$ can then be parametrized as:
\begin{align*}
\bm{\sigma}_{12}(s) &= \tanh\frac{s(1-s)}{\epsilon}\,\x\,,\\[5pt]
\bm{\sigma}_{23}(s) &= \tanh\frac{s(1-s)}{\epsilon}\,\left(-\tfrac{1}{2}\,\x+\tfrac{\sqrt{3}}{2}\,\y\right)\,,\\[5pt]
\bm{\sigma}_{31}(s) &= \tanh\frac{s(1-s)}{\epsilon}\,\left(\tfrac{1}{2}\,\x+\tfrac{\sqrt{3}}{2}\,\y\right)\,,
\end{align*}
where $\epsilon$ in this case is defined as  $\epsilon^{2}=2K_{A}/\lambda t_{0}^{2}\ell^{2}$ with $\ell$ the edge length of the octahedron.
The tilt vector resulting from this model of the crystallographic configuration is sketched in Fig. \ref{fig:textures2} (right).

\begin{figure}
\centering
\includegraphics[width=0.7\columnwidth]{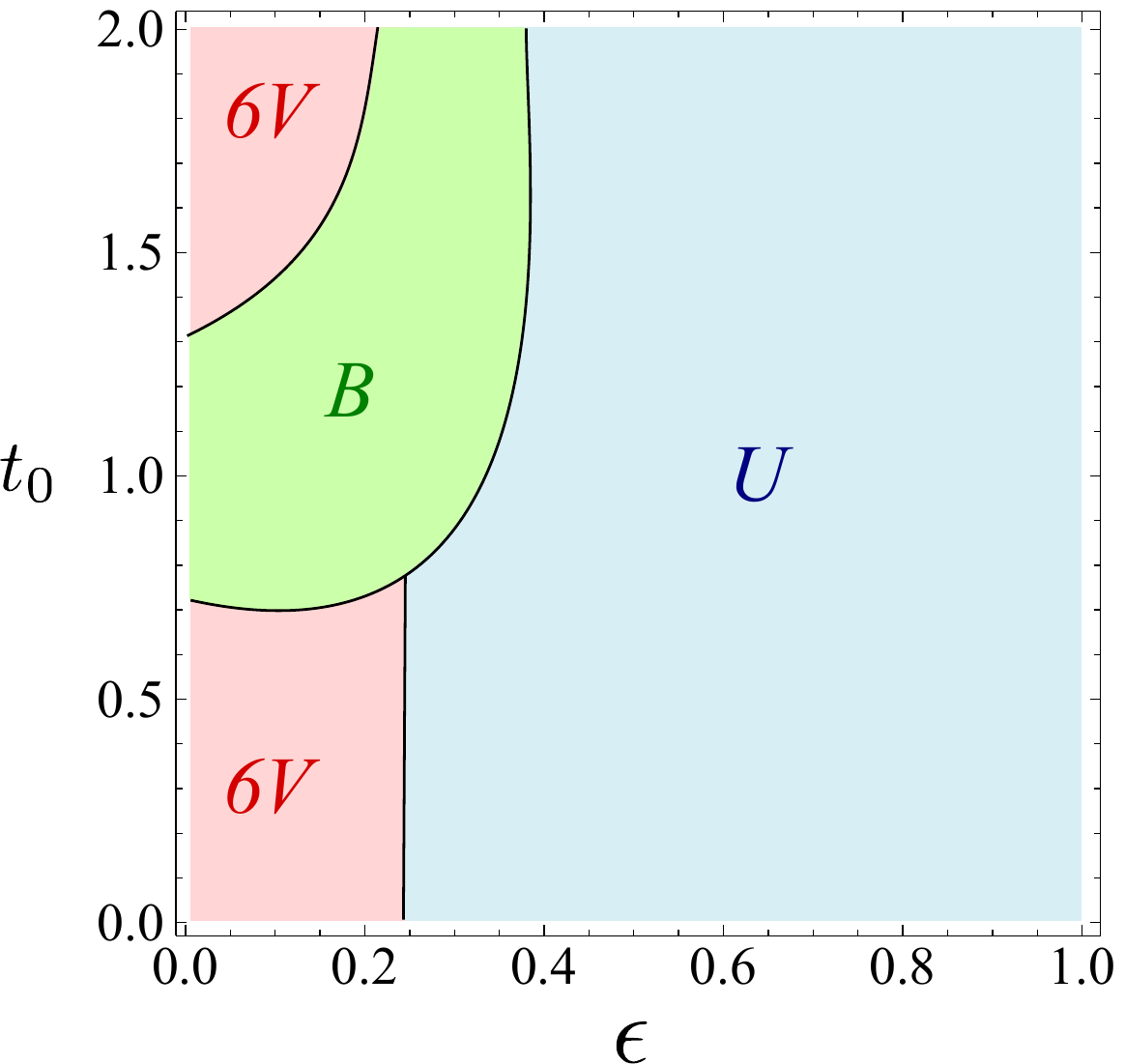}
\caption{\label{fig:phase-diagram}Phase-diagram of the various configurations discussed in the text on an octahedral nanoparticle. $U$ represents the untilted configuration in which the direction ${\bf n}$ of the chains is everywhere parallel to the normal vector ${\bf N}$, thus ${\bf t}={\bf 0}$. {\em 6V} is the 6-valent configuration associated with the crystallographic model and {\em B} is the octahedral analog of the bipolar texture.}
\end{figure}

The octahedral analog of the bipolar configuration described in the previous section, gives rise to a situation in between the continuous and the crystallographic model. Using again as a reference the triangle on the positive octant of the lab $\{\X,\,\Y,\,\Z\}$ frame, we have that:
\[
{\bf n}_{1} = \X\,,\qquad
{\bf n}_{2} = \Y\,,\qquad
{\bf n}_{3} = {\bf N}\,.
\]
Thus the vector field has a source (sink) at the apex of the octahedron, where the chains are parallel to the normal vector ${\bf N}$, and rotates smoothly across the equator where the two square pyramids forming the octahedron join. As for the continuous model, we can construct the tilt vector along the entire perimeter of a triangular face by linearly interpolating between the three vectors ${\bf n}_{1}$, ${\bf n}_{2}$ and ${\bf n}_{3}$. This gives us:
\begin{align*}
{\bf n}_{12}(s) &= (1-s)\X+s\Y\,,\\[5pt]
{\bf n}_{23}(s) &= (1-s)\Y+s{\bf N}\,,\\[5pt]
{\bf n}_{31}(s) &= (1-s){\bf N}+s\X\,.
\end{align*}
Proceeding as in the previous case, one can construct the function $\bm{\sigma}$ as shown below:
\begin{align*}
\bm{\sigma}_{12}(s) &= t_{0}^{-1}\left[\sqrt{\tfrac{3}{2}}\,(2s-1)\,\x-\sqrt{\tfrac{1}{2}}\,\y\right],\\[5pt]
\bm{\sigma}_{23}(s) &= \frac{s-1}{t_{0}[1+(\sqrt{3}-1)s]}\,\left(-\sqrt{\tfrac{3}{2}}\,\x+\sqrt{\tfrac{1}{2}}\,\y\right)\,,\\[5pt]
\bm{\sigma}_{31}(s) &= \frac{s}{t_{0}[(\sqrt{3}-1)s+\sqrt{3}]}\,\left(\sqrt{\tfrac{3}{2}}\,\x+\sqrt{\tfrac{1}{2}}\,\y\right)\,.
\end{align*}
With the function $\bm{\sigma}$ at hand, we can solve the boundary-value problem \eqref{eq:landau-ginzburg2} by numerical integration. This was performed for different values of $\epsilon$ and $t_{0}$ using a commercial finite element software and is summarized in the phase-diagram of Fig. \ref{fig:phase-diagram}, where $U$ represent the untilted configuration, {\em 6V} is the 6-valent configuration associated with the crystallographic model and {\em B} is the octahedral analog of the bipolar texture. As expected, the defective configurations are replaced by the untilted one at large epsilon, where the energetic cost of spatial variations in the chains tilt becomes too expensive. For small $\epsilon$ both the bipolar and the $6$-valent configuration are stable depending on the value of the preferential tilt $t_{0}$. The smoothly rotating configuration associated with the continuous model, on the other hand, is never an
energy minimizer due to its highly defective structure. Such a configuration, however, is likely to occur at finite temperature where entropy favors the proliferation of defects.

As noted in the introduction, defects are not only mathematically and physically distinguished sites, but also they can be chemically functionalized via place-exchange reactions. Polymerization chemistry then allows the functionalized NPs to be linked into chains and one can explore polymer physics with polymers based on NP monomers. Linear chains of NPs ranging from $3$ to $20$ monomers have been observed~\cite{DeVries:2007}. Because the local tilt geometry is different for distinct types of defects, they are likely to have different chemical reactivity.
{\em 6V} textures with 4 additional defects available for functionalizing provide novel possibilities for molecules and bulk materials including 2D films, 3D crystal structures and branched structures. Observation of any of these higher-coordination number structures would provide evidence for the {\em 6V} texture. Detailed analyses of relative yields of various structures could then provide bounds on the relative chemical reactivity of different types of defect~\cite{stellacci:2008}.

{\bf Conclusion.} $-$ In this article we have investigated the energetically-preferred configurations adopted by tilted chain-like molecules coating a metal nanoparticle by analyzing a Ginzburg-Landau model for the system. Spherical nanoparticles have a low temperature ordered phase with a bipolar texture characterized by index one defects at the poles. Functionalization of such particles together with linking gives rise to polymer chains. For small nanoparticles the faceted shape of the particle may be significant. We extend our analysis to Ginzburg-Landau models on octahedra and determine the phase diagram as a function of the equilibrium tilt angle $t_0$ and a dimensionless parameter $\epsilon$ characterizing the relative strength of the van der Waals interactions between chains and the stiffness modulus $K_A$ controlling spatial inhomogeneities in the tilt.  We recover an octahedral generalization of the bipolar texture but also a qualitatively new texture with 6 point defects ({\em 6V}), favored at strong van der Waals couplings. Thus shape can dramatically alter the ground state and favor extra defects. {\em 6V} textures allow for the creation of valence 6 NPs after functionalization of the defects and consequently new classes of molecules and bulk materials when nanoparticles are linked.

\begin{center}
$***$
\end{center}

LG and MJB would like to thank Francesco Stellacci for the insightful conversations that inspired this work. We gratefully acknowledge support from the Wyss Institute (LG), the Harvard Kavli Institue for Bionano Science \& Technology (LG), the NSF Harvard MRSEC (LG). The work of MJB and XM was supported by the National Science Foundation grant DMR-0808812. AM is supported by Award No. KUK-C1-013-04, made by King Abdullah University of Science and Technology (KAUST), to the Oxford Centre for Collaborative Applied Mathematics and an EPSRC Career Acceleration Fellowship EP/J001686/1.

\bibliography{biblio}

\begin{thebibliography}{28}%
\makeatletter
\providecommand \@ifxundefined [1]{%
 \@ifx{#1\undefined}
}%
\providecommand \@ifnum [1]{%
 \ifnum #1\expandafter \@firstoftwo
 \else \expandafter \@secondoftwo
 \fi
}%
\providecommand \@ifx [1]{%
 \ifx #1\expandafter \@firstoftwo
 \else \expandafter \@secondoftwo
 \fi
}%
\providecommand \natexlab [1]{#1}%
\providecommand \enquote  [1]{``#1''}%
\providecommand \bibnamefont  [1]{#1}%
\providecommand \bibfnamefont [1]{#1}%
\providecommand \citenamefont [1]{#1}%
\providecommand \href@noop [0]{\@secondoftwo}%
\providecommand \href [0]{\begingroup \@sanitize@url \@href}%
\providecommand \@href[1]{\@@startlink{#1}\@@href}%
\providecommand \@@href[1]{\endgroup#1\@@endlink}%
\providecommand \@sanitize@url [0]{\catcode `\\12\catcode `\$12\catcode
  `\&12\catcode `\#12\catcode `\^12\catcode `\_12\catcode `\%12\relax}%
\providecommand \@@startlink[1]{}%
\providecommand \@@endlink[0]{}%
\providecommand \url  [0]{\begingroup\@sanitize@url \@url }%
\providecommand \@url [1]{\endgroup\@href {#1}{\urlprefix }}%
\providecommand \urlprefix  [0]{URL }%
\providecommand \Eprint [0]{\href }%
\providecommand \doibase [0]{http://dx.doi.org/}%
\providecommand \selectlanguage [0]{\@gobble}%
\providecommand \bibinfo  [0]{\@secondoftwo}%
\providecommand \bibfield  [0]{\@secondoftwo}%
\providecommand \translation [1]{[#1]}%
\providecommand \BibitemOpen [0]{}%
\providecommand \bibitemStop [0]{}%
\providecommand \bibitemNoStop [0]{.\EOS\space}%
\providecommand \EOS [0]{\spacefactor3000\relax}%
\providecommand \BibitemShut  [1]{\csname bibitem#1\endcsname}%
\let\auto@bib@innerbib\@empty
%</preamble>
\bibitem [{\citenamefont {Daniel}\ and\ \citenamefont
  {Astruc}(2004)}]{Daniel:2004}%
  \BibitemOpen
  \bibfield  {author} {\bibinfo {author} {\bibfnamefont {M.-C.}\ \bibnamefont
  {Daniel}}\ and\ \bibinfo {author} {\bibfnamefont {D.}~\bibnamefont
  {Astruc}},\ }\href@noop {} {\bibfield  {journal} {\bibinfo  {journal} {Chem.
  Rev.}\ }\textbf {\bibinfo {volume} {104}},\ \bibinfo {pages} {293} (\bibinfo
  {year} {2004})}\BibitemShut {NoStop}%
\bibitem [{\citenamefont {Love}\ \emph {et~al.}(2005)\citenamefont {Love},
  \citenamefont {Estroff}, \citenamefont {Kriebel}, \citenamefont {Nuzzo},\
  and\ \citenamefont {Whitesides}}]{Love:2005}%
  \BibitemOpen
  \bibfield  {author} {\bibinfo {author} {\bibfnamefont {J.~C.}\ \bibnamefont
  {Love}}, \bibinfo {author} {\bibfnamefont {L.~A.}\ \bibnamefont {Estroff}},
  \bibinfo {author} {\bibfnamefont {J.~K.}\ \bibnamefont {Kriebel}}, \bibinfo
  {author} {\bibfnamefont {R.~G.}\ \bibnamefont {Nuzzo}}, \ and\ \bibinfo
  {author} {\bibfnamefont {G.~M.}\ \bibnamefont {Whitesides}},\ }\href@noop {}
  {\bibfield  {journal} {\bibinfo  {journal} {Chem. Rev.}\ }\textbf {\bibinfo
  {volume} {105}},\ \bibinfo {pages} {1103} (\bibinfo {year}
  {2005})}\BibitemShut {NoStop}%
\bibitem [{\citenamefont {Andres}\ \emph {et~al.}(1996)\citenamefont {Andres},
  \citenamefont {Bielefeld}, \citenamefont {Henderson}, \citenamefont {Janes},
  \citenamefont {Kolagunta}, \citenamefont {Kubiak}, \citenamefont {Mahoney},\
  and\ \citenamefont {Osifchin}}]{andres:1996}%
  \BibitemOpen
  \bibfield  {author} {\bibinfo {author} {\bibfnamefont {R.~P.}\ \bibnamefont
  {Andres}}, \bibinfo {author} {\bibfnamefont {J.~D.}\ \bibnamefont
  {Bielefeld}}, \bibinfo {author} {\bibfnamefont {J.~I.}\ \bibnamefont
  {Henderson}}, \bibinfo {author} {\bibfnamefont {D.~B.}\ \bibnamefont
  {Janes}}, \bibinfo {author} {\bibfnamefont {V.~R.}\ \bibnamefont
  {Kolagunta}}, \bibinfo {author} {\bibfnamefont {C.~P.}\ \bibnamefont
  {Kubiak}}, \bibinfo {author} {\bibfnamefont {W.~J.}\ \bibnamefont {Mahoney}},
  \ and\ \bibinfo {author} {\bibfnamefont {R.~G.}\ \bibnamefont {Osifchin}},\
  }\href@noop {} {\bibfield  {journal} {\bibinfo  {journal} {Science}\ }\textbf
  {\bibinfo {volume} {273}},\ \bibinfo {pages} {1690} (\bibinfo {year}
  {1996})}\BibitemShut {NoStop}%
\bibitem [{\citenamefont {Collier}\ \emph {et~al.}(1997)\citenamefont
  {Collier}, \citenamefont {Saykally}, \citenamefont {Shiang}, \citenamefont
  {Henrichs},\ and\ \citenamefont {Heath}}]{collier:1997}%
  \BibitemOpen
  \bibfield  {author} {\bibinfo {author} {\bibfnamefont {C.~P.}\ \bibnamefont
  {Collier}}, \bibinfo {author} {\bibfnamefont {R.~J.}\ \bibnamefont
  {Saykally}}, \bibinfo {author} {\bibfnamefont {J.~J.}\ \bibnamefont
  {Shiang}}, \bibinfo {author} {\bibfnamefont {S.~E.}\ \bibnamefont
  {Henrichs}}, \ and\ \bibinfo {author} {\bibfnamefont {J.~R.}\ \bibnamefont
  {Heath}},\ }\href@noop {} {\bibfield  {journal} {\bibinfo  {journal}
  {Science}\ }\textbf {\bibinfo {volume} {277}},\ \bibinfo {pages} {1978}
  (\bibinfo {year} {1997})}\BibitemShut {NoStop}%
\bibitem [{\citenamefont {Han}\ \emph {et~al.}(2007)\citenamefont {Han},
  \citenamefont {Ghosh},\ and\ \citenamefont {Rotello}}]{han:2007}%
  \BibitemOpen
  \bibfield  {author} {\bibinfo {author} {\bibfnamefont {G.}~\bibnamefont
  {Han}}, \bibinfo {author} {\bibfnamefont {P.}~\bibnamefont {Ghosh}}, \ and\
  \bibinfo {author} {\bibfnamefont {V.~M.}\ \bibnamefont {Rotello}},\
  }\href@noop {} {\bibfield  {journal} {\bibinfo  {journal} {Nanomedicine}\
  }\textbf {\bibinfo {volume} {2}},\ \bibinfo {pages} {113} (\bibinfo {year}
  {2007})}\BibitemShut {NoStop}%
\bibitem [{\citenamefont {Pasquato}\ \emph {et~al.}(2004)\citenamefont
  {Pasquato}, \citenamefont {Pengo},\ and\ \citenamefont
  {Scrimin}}]{pasquato:2005}%
  \BibitemOpen
  \bibfield  {author} {\bibinfo {author} {\bibfnamefont {L.}~\bibnamefont
  {Pasquato}}, \bibinfo {author} {\bibfnamefont {P.}~\bibnamefont {Pengo}}, \
  and\ \bibinfo {author} {\bibfnamefont {P.}~\bibnamefont {Scrimin}},\
  }\href@noop {} {\bibfield  {journal} {\bibinfo  {journal} {Supramol. Chem.}\
  }\textbf {\bibinfo {volume} {17}},\ \bibinfo {pages} {163} (\bibinfo {year}
  {2004})}\BibitemShut {NoStop}%
\bibitem [{\citenamefont {Georganopoulou}\ \emph {et~al.}(2005)\citenamefont
  {Georganopoulou}, \citenamefont {Chang}, \citenamefont {Naam}, \citenamefont
  {Thaxton}, \citenamefont {Mufson}, \citenamefont {Klein},\ and\ \citenamefont
  {Markin}}]{georganopoulou:2005}%
  \BibitemOpen
  \bibfield  {author} {\bibinfo {author} {\bibfnamefont {D.~G.}\ \bibnamefont
  {Georganopoulou}}, \bibinfo {author} {\bibfnamefont {L.}~\bibnamefont
  {Chang}}, \bibinfo {author} {\bibfnamefont {J.~M.}\ \bibnamefont {Naam}},
  \bibinfo {author} {\bibfnamefont {C.~S.}\ \bibnamefont {Thaxton}}, \bibinfo
  {author} {\bibfnamefont {E.~J.}\ \bibnamefont {Mufson}}, \bibinfo {author}
  {\bibfnamefont {W.~L.}\ \bibnamefont {Klein}}, \ and\ \bibinfo {author}
  {\bibfnamefont {C.~A.}\ \bibnamefont {Markin}},\ }\href@noop {} {\bibfield
  {journal} {\bibinfo  {journal} {Proc. Natl. Acad. Sci. U.S.A.}\ }\textbf
  {\bibinfo {volume} {102}},\ \bibinfo {pages} {2273} (\bibinfo {year}
  {2005})}\BibitemShut {NoStop}%
\bibitem [{\citenamefont {Ditlbacher}\ \emph {et~al.}(2002)\citenamefont
  {Ditlbacher}, \citenamefont {Krenn}, \citenamefont {Schider}, \citenamefont
  {Leitner},\ and\ \citenamefont {Aussenberg}}]{Ditlbacher:2002}%
  \BibitemOpen
  \bibfield  {author} {\bibinfo {author} {\bibfnamefont {H.}~\bibnamefont
  {Ditlbacher}}, \bibinfo {author} {\bibfnamefont {J.}~\bibnamefont {Krenn}},
  \bibinfo {author} {\bibfnamefont {G.}~\bibnamefont {Schider}}, \bibinfo
  {author} {\bibfnamefont {A.}~\bibnamefont {Leitner}}, \ and\ \bibinfo
  {author} {\bibfnamefont {F.}~\bibnamefont {Aussenberg}},\ }\href@noop {}
  {\bibfield  {journal} {\bibinfo  {journal} {Appl. Phys. Lett.}\ }\textbf
  {\bibinfo {volume} {81}},\ \bibinfo {pages} {1762} (\bibinfo {year}
  {2002})}\BibitemShut {NoStop}%
\bibitem [{\citenamefont {MacFarlane}\ \emph {et~al.}(2011)\citenamefont
  {MacFarlane}, \citenamefont {Lee}, \citenamefont {Jones}, \citenamefont
  {Harris}, \citenamefont {Schatz},\ and\ \citenamefont
  {Mirkin}}]{Mirkin:2011}%
  \BibitemOpen
  \bibfield  {author} {\bibinfo {author} {\bibfnamefont {R.}~\bibnamefont
  {MacFarlane}}, \bibinfo {author} {\bibfnamefont {B.}~\bibnamefont {Lee}},
  \bibinfo {author} {\bibfnamefont {M.}~\bibnamefont {Jones}}, \bibinfo
  {author} {\bibfnamefont {N.}~\bibnamefont {Harris}}, \bibinfo {author}
  {\bibfnamefont {G.}~\bibnamefont {Schatz}}, \ and\ \bibinfo {author}
  {\bibfnamefont {C.}~\bibnamefont {Mirkin}},\ }\href@noop {} {\bibfield
  {journal} {\bibinfo  {journal} {Science}\ }\textbf {\bibinfo {volume}
  {334}},\ \bibinfo {pages} {204} (\bibinfo {year} {2011})}\BibitemShut
  {NoStop}%
\bibitem [{\citenamefont {Salata}(2004)}]{Salata:2004}%
  \BibitemOpen
  \bibfield  {author} {\bibinfo {author} {\bibfnamefont {O.}~\bibnamefont
  {Salata}},\ }\href@noop {} {\bibfield  {journal} {\bibinfo  {journal} {J.
  Nanobiotechnology}\ }\textbf {\bibinfo {volume} {2}},\ \bibinfo {pages} {3}
  (\bibinfo {year} {2004})}\BibitemShut {NoStop}%
\bibitem [{\citenamefont {Luedtke}\ and\ \citenamefont
  {Landman}(1998)}]{Landman:1998}%
  \BibitemOpen
  \bibfield  {author} {\bibinfo {author} {\bibfnamefont {W.~D.}\ \bibnamefont
  {Luedtke}}\ and\ \bibinfo {author} {\bibfnamefont {U.}~\bibnamefont
  {Landman}},\ }\href@noop {} {\bibfield  {journal} {\bibinfo  {journal} {J.
  Phys. Chem. B}\ }\textbf {\bibinfo {volume} {102}},\ \bibinfo {pages} {6566}
  (\bibinfo {year} {1998})}\BibitemShut {NoStop}%
\bibitem [{\citenamefont {Zanchet}\ \emph {et~al.}(2000)\citenamefont
  {Zanchet}, \citenamefont {Hall},\ and\ \citenamefont
  {Ugarte}}]{Zanchet:2000}%
  \BibitemOpen
  \bibfield  {author} {\bibinfo {author} {\bibfnamefont {D.}~\bibnamefont
  {Zanchet}}, \bibinfo {author} {\bibfnamefont {B.~D.}\ \bibnamefont {Hall}}, \
  and\ \bibinfo {author} {\bibfnamefont {D.~J.}\ \bibnamefont {Ugarte}},\
  }\href@noop {} {\bibfield  {journal} {\bibinfo  {journal} {Phys. Chem. B}\
  }\textbf {\bibinfo {volume} {104}},\ \bibinfo {pages} {11013} (\bibinfo
  {year} {2000})}\BibitemShut {NoStop}%
\bibitem [{\citenamefont {Landman}\ and\ \citenamefont
  {Luedtke}(2004)}]{Landman:2004}%
  \BibitemOpen
  \bibfield  {author} {\bibinfo {author} {\bibfnamefont {U.}~\bibnamefont
  {Landman}}\ and\ \bibinfo {author} {\bibfnamefont {W.~D.}\ \bibnamefont
  {Luedtke}},\ }\href@noop {} {\bibfield  {journal} {\bibinfo  {journal}
  {Faraday Discuss.}\ }\textbf {\bibinfo {volume} {125}},\ \bibinfo {pages} {1}
  (\bibinfo {year} {2004})}\BibitemShut {NoStop}%
\bibitem [{\citenamefont {Ghorai}\ and\ \citenamefont
  {Glotzer}(2007)}]{Ghorai:2007}%
  \BibitemOpen
  \bibfield  {author} {\bibinfo {author} {\bibfnamefont {P.~K.}\ \bibnamefont
  {Ghorai}}\ and\ \bibinfo {author} {\bibfnamefont {S.}~\bibnamefont
  {Glotzer}},\ }\href@noop {} {\bibfield  {journal} {\bibinfo  {journal} {J.
  Phys. Chem. C}\ }\textbf {\bibinfo {volume} {111}},\ \bibinfo {pages} {15857}
  (\bibinfo {year} {2007})}\BibitemShut {NoStop}%
\bibitem [{\citenamefont {Eisenberg}\ and\ \citenamefont
  {Guy}(1979)}]{Eisenberg:1979}%
  \BibitemOpen
  \bibfield  {author} {\bibinfo {author} {\bibfnamefont {M.}~\bibnamefont
  {Eisenberg}}\ and\ \bibinfo {author} {\bibfnamefont {R.}~\bibnamefont
  {Guy}},\ }\href@noop {} {\bibfield  {journal} {\bibinfo  {journal} {Am. Math.
  Mon.}\ }\textbf {\bibinfo {volume} {86}},\ \bibinfo {pages} {571} (\bibinfo
  {year} {1979})}\BibitemShut {NoStop}%
\bibitem [{\citenamefont {Bowick}\ and\ \citenamefont
  {Giomi}(2009)}]{Bowick:2009}%
  \BibitemOpen
  \bibfield  {author} {\bibinfo {author} {\bibfnamefont {M.~J.}\ \bibnamefont
  {Bowick}}\ and\ \bibinfo {author} {\bibfnamefont {L.}~\bibnamefont {Giomi}},\
  }\href@noop {} {\bibfield  {journal} {\bibinfo  {journal} {Adv. Phys.}\
  }\textbf {\bibinfo {volume} {58}},\ \bibinfo {pages} {449} (\bibinfo {year}
  {2009})}\BibitemShut {NoStop}%
\bibitem [{\citenamefont {Turner}\ \emph {et~al.}(2010)\citenamefont {Turner},
  \citenamefont {Vitelli},\ and\ \citenamefont {Nelson}}]{Turner:2010}%
  \BibitemOpen
  \bibfield  {author} {\bibinfo {author} {\bibfnamefont {A.~M.}\ \bibnamefont
  {Turner}}, \bibinfo {author} {\bibfnamefont {V.}~\bibnamefont {Vitelli}}, \
  and\ \bibinfo {author} {\bibfnamefont {D.~R.}\ \bibnamefont {Nelson}},\
  }\href@noop {} {\bibfield  {journal} {\bibinfo  {journal} {Rev. Mod. Phys.}\
  }\textbf {\bibinfo {volume} {82}},\ \bibinfo {pages} {1301} (\bibinfo {year}
  {2010})}\BibitemShut {NoStop}%
\bibitem [{\citenamefont {Nelson}(2002)}]{Nelson:2002}%
  \BibitemOpen
  \bibfield  {author} {\bibinfo {author} {\bibfnamefont {D.~R.}\ \bibnamefont
  {Nelson}},\ }\href@noop {} {\bibfield  {journal} {\bibinfo  {journal} {Nano
  Lett.}\ }\textbf {\bibinfo {volume} {2}},\ \bibinfo {pages} {1125} (\bibinfo
  {year} {2002})}\BibitemShut {NoStop}%
\bibitem [{\citenamefont {DeVries}\ \emph {et~al.}(2007)\citenamefont
  {DeVries}, \citenamefont {Brunnbauer}, \citenamefont {Hu}, \citenamefont
  {Jackson}, \citenamefont {Long}, \citenamefont {Neltner}, \citenamefont
  {Uzun}, \citenamefont {Wunsch},\ and\ \citenamefont
  {Stellacci}}]{DeVries:2007}%
  \BibitemOpen
  \bibfield  {author} {\bibinfo {author} {\bibfnamefont {G.~A.}\ \bibnamefont
  {DeVries}}, \bibinfo {author} {\bibfnamefont {M.}~\bibnamefont {Brunnbauer}},
  \bibinfo {author} {\bibfnamefont {Y.}~\bibnamefont {Hu}}, \bibinfo {author}
  {\bibfnamefont {A.~M.}\ \bibnamefont {Jackson}}, \bibinfo {author}
  {\bibfnamefont {B.}~\bibnamefont {Long}}, \bibinfo {author} {\bibfnamefont
  {B.~T.}\ \bibnamefont {Neltner}}, \bibinfo {author} {\bibfnamefont
  {O.}~\bibnamefont {Uzun}}, \bibinfo {author} {\bibfnamefont {B.~H.}\
  \bibnamefont {Wunsch}}, \ and\ \bibinfo {author} {\bibfnamefont
  {F.}~\bibnamefont {Stellacci}},\ }\href@noop {} {\bibfield  {journal}
  {\bibinfo  {journal} {Science}\ }\textbf {\bibinfo {volume} {315}},\ \bibinfo
  {pages} {358} (\bibinfo {year} {2007})}\BibitemShut {NoStop}%
\bibitem [{\citenamefont {Rapino}\ and\ \citenamefont
  {Zerbetto}(2007)}]{Zerbetto:2007}%
  \BibitemOpen
  \bibfield  {author} {\bibinfo {author} {\bibfnamefont {S.}~\bibnamefont
  {Rapino}}\ and\ \bibinfo {author} {\bibfnamefont {F.}~\bibnamefont
  {Zerbetto}},\ }\href@noop {} {\bibfield  {journal} {\bibinfo  {journal}
  {Small}\ }\textbf {\bibinfo {volume} {3}},\ \bibinfo {pages} {386} (\bibinfo
  {year} {2007})}\BibitemShut {NoStop}%
\bibitem [{\citenamefont {MacKintosh}\ and\ \citenamefont
  {Lubensky}(1991)}]{MacKintosh:1991}%
  \BibitemOpen
  \bibfield  {author} {\bibinfo {author} {\bibfnamefont {F.~C.}\ \bibnamefont
  {MacKintosh}}\ and\ \bibinfo {author} {\bibfnamefont {T.~C.}\ \bibnamefont
  {Lubensky}},\ }\href@noop {} {\bibfield  {journal} {\bibinfo  {journal}
  {Phys. Rev. Lett.}\ }\textbf {\bibinfo {volume} {67}},\ \bibinfo {pages}
  {1169} (\bibinfo {year} {1991})}\BibitemShut {NoStop}%
\bibitem [{\citenamefont {Outka}\ \emph {et~al.}(1987)\citenamefont {Outka},
  \citenamefont {St{\"o}hr}, \citenamefont {Rabe}, \citenamefont {Swalen},\
  and\ \citenamefont {Rotermund}}]{Outka:1987}%
  \BibitemOpen
  \bibfield  {author} {\bibinfo {author} {\bibfnamefont {D.~A.}\ \bibnamefont
  {Outka}}, \bibinfo {author} {\bibfnamefont {J.}~\bibnamefont {St{\"o}hr}},
  \bibinfo {author} {\bibfnamefont {J.}~\bibnamefont {Rabe}}, \bibinfo {author}
  {\bibfnamefont {J.~D.}\ \bibnamefont {Swalen}}, \ and\ \bibinfo {author}
  {\bibfnamefont {H.~H.}\ \bibnamefont {Rotermund}},\ }\href@noop {} {\bibfield
   {journal} {\bibinfo  {journal} {Phys. Rev. Lett.}\ }\textbf {\bibinfo
  {volume} {59}},\ \bibinfo {pages} {1321} (\bibinfo {year}
  {1987})}\BibitemShut {NoStop}%
\bibitem [{\citenamefont {Sellers}\ \emph {et~al.}(1993)\citenamefont
  {Sellers}, \citenamefont {Ulman}, \citenamefont {Shnidman},\ and\
  \citenamefont {Eilers}}]{sellers:1993}%
  \BibitemOpen
  \bibfield  {author} {\bibinfo {author} {\bibfnamefont {H.}~\bibnamefont
  {Sellers}}, \bibinfo {author} {\bibfnamefont {A.}~\bibnamefont {Ulman}},
  \bibinfo {author} {\bibfnamefont {Y.}~\bibnamefont {Shnidman}}, \ and\
  \bibinfo {author} {\bibfnamefont {J.~E.}\ \bibnamefont {Eilers}},\
  }\href@noop {} {\bibfield  {journal} {\bibinfo  {journal} {J. Am. Chem.
  Soc.}\ }\textbf {\bibinfo {volume} {115}},\ \bibinfo {pages} {9389} (\bibinfo
  {year} {1993})}\BibitemShut {NoStop}%
\bibitem [{\citenamefont {Porter}\ \emph {et~al.}(1987)\citenamefont {Porter},
  \citenamefont {Bright}, \citenamefont {Allara},\ and\ \citenamefont
  {Chidsey}}]{porter:1987}%
  \BibitemOpen
  \bibfield  {author} {\bibinfo {author} {\bibfnamefont {M.~D.}\ \bibnamefont
  {Porter}}, \bibinfo {author} {\bibfnamefont {T.~B.}\ \bibnamefont {Bright}},
  \bibinfo {author} {\bibfnamefont {D.~L.}\ \bibnamefont {Allara}}, \ and\
  \bibinfo {author} {\bibfnamefont {C.~E.~D.}\ \bibnamefont {Chidsey}},\
  }\href@noop {} {\bibfield  {journal} {\bibinfo  {journal} {J. Am. Chem.
  Soc.}\ }\textbf {\bibinfo {volume} {109}},\ \bibinfo {pages} {3559} (\bibinfo
  {year} {1987})}\BibitemShut {NoStop}%
\bibitem [{\citenamefont {Nuzzo}\ \emph {et~al.}(1990)\citenamefont {Nuzzo},
  \citenamefont {Dubois},\ and\ \citenamefont {Allara}}]{nuzzo:1990}%
  \BibitemOpen
  \bibfield  {author} {\bibinfo {author} {\bibfnamefont {R.~G.}\ \bibnamefont
  {Nuzzo}}, \bibinfo {author} {\bibfnamefont {L.~H.}\ \bibnamefont {Dubois}}, \
  and\ \bibinfo {author} {\bibfnamefont {D.~L.}\ \bibnamefont {Allara}},\
  }\href@noop {} {\bibfield  {journal} {\bibinfo  {journal} {J. Am. Chem.
  Soc.}\ }\textbf {\bibinfo {volume} {112}},\ \bibinfo {pages} {558} (\bibinfo
  {year} {1990})}\BibitemShut {NoStop}%
\bibitem [{\citenamefont {Ehler}\ \emph {et~al.}(1997)\citenamefont {Ehler},
  \citenamefont {Malmberg},\ and\ \citenamefont {Noe}}]{ehler:1997}%
  \BibitemOpen
  \bibfield  {author} {\bibinfo {author} {\bibfnamefont {T.~T.}\ \bibnamefont
  {Ehler}}, \bibinfo {author} {\bibfnamefont {N.}~\bibnamefont {Malmberg}}, \
  and\ \bibinfo {author} {\bibfnamefont {L.~J.}\ \bibnamefont {Noe}},\
  }\href@noop {} {\bibfield  {journal} {\bibinfo  {journal} {J. Phys. Chem. B}\
  }\textbf {\bibinfo {volume} {101}},\ \bibinfo {pages} {1268} (\bibinfo {year}
  {1997})}\BibitemShut {NoStop}%
\bibitem [{\citenamefont {Tinkham}(1996)}]{Tinkham}%
  \BibitemOpen
  \bibfield  {author} {\bibinfo {author} {\bibfnamefont {M.}~\bibnamefont
  {Tinkham}},\ }\href@noop {} {\emph {\bibinfo {title} {Introduction to
  Superconductivity}}}\ (\bibinfo  {publisher} {McGraw-Hill},\ \bibinfo
  {address} {New York},\ \bibinfo {year} {1996})\BibitemShut {NoStop}%
\bibitem [{\citenamefont {DeVries}\ \emph {et~al.}(2008)\citenamefont
  {DeVries}, \citenamefont {Talley}, \citenamefont {Carney},\ and\
  \citenamefont {Stellacci}}]{stellacci:2008}%
  \BibitemOpen
  \bibfield  {author} {\bibinfo {author} {\bibfnamefont {G.~A.}\ \bibnamefont
  {DeVries}}, \bibinfo {author} {\bibfnamefont {F.~R.}\ \bibnamefont {Talley}},
  \bibinfo {author} {\bibfnamefont {R.~P.}\ \bibnamefont {Carney}}, \ and\
  \bibinfo {author} {\bibfnamefont {F.}~\bibnamefont {Stellacci}},\ }\href@noop
  {} {\bibfield  {journal} {\bibinfo  {journal} {Adv. Mater.}\ }\textbf
  {\bibinfo {volume} {20}},\ \bibinfo {pages} {4243} (\bibinfo {year}
  {2008})}\BibitemShut {NoStop}%
\end{thebibliography}%
\end{document}